# Electric potential of a metallic nanowall between cathode and anode planes

Xi-Zhou Qin, Wie-Liang Wang, Zhi-Bing Li∗

State Key Laboratory of Optoelectronic Materials and Technologies

School of Physics and Engineering, Sun Yat-Sen University, Guangzhou 510275, P.R. China

**Abstract**

We obtained the exact expression of the electric potential in the space around a nanowall that is vertically mounted on a planar cathode. The system is designed as a cold field electron emitter or an electron tunneling line scanner. The finite cathode-anode distance has been taking into account. The analytical results are compared with that obtained by the finite- element method.

Owing to the nanotechnology, the cold field electron emission (CFE) has become or been close to a practical microelectronic vacuum electron source that may be used in flat-panel displays, electron microscope and in parallel e-beam lithography systems. The development of nanotechnology also depends heavily on the atomic scale imaging/manipulation, for instances, achieved via electron (homographic) microscopy, field emission microscopy, scanning tunneling microscopy (STM), and atomic force microscopy. Usually, nano-tips are used as the emitter. A nature reason is that the high aspect ratio of nano-tips can enhance the apex field a great deal such that the field electron emission can be driven by macroscopic electric fields of about ten volts per micrometer or less. This has been demonstrated by the Spindt-type cathodes, which is basically micro-fabricated molybdenum tips in gated configuration [1]. In recent years, much interest has turned to the nano-structures, such as the carbon nanotubes and nanowires of various materials [2, 3], of which the aspect ratio can be easily a few thousands. However, for some usages, the electron beam produced by nanotubes and nanowires would be too bright and too concentrated. It requires superb skill to control the uniformity of field emission in large area.

---

∗ Corresponding author: stslzb@mail.sysu.edu.cn

Recently, two-dimensional field emitters, particularly the graphene [4-5], have attracted considerable attention. Several groups have demonstrated that graphene does show promising CFE properties, such as a low emission threshold field and large emission current density [6-12]. Theoretically, it has been shown that the CFE from two-dimensional structures would have a current-field characteristic that is completely different from the conventional Fowler-Nordheim (FN) law [14]. The conventional FN theory for the characteristic relation between the CFE current and the applied macroscopic electric field was derived for the planar emitters in principle [FN]. For two-dimensional nanowalls, two new features should be considered. The first is the electron supply function of two-dimensional electron system that has density of states different from that of the three-dimensional systems. Recently, an investigation on CFE of the graphene that takes into account the band structure has been done [Wang preprint]. The second is the electric potential in the vacuum gap between the cathode and the anode. An analytical expression of the electric potential for nanowalls has been obtained in the case that the cathode-anode distance (CAD) is remote [14].

The present paper should study the finite cathode-anode distance effect. That is important for the application in STM and in the close field imaging, for instances. It is found that the finite CAD correction is large even when the ratio of CAD over the height of nanowall reaches several decades. The solution will be compared with the finite element simulation (FES).

**1. The model**

Consider an infinite nanowall mounted on a planar cathode vertically. The cross section of the set up is shown in Fig.1, where the bottom and the top planes are the cathode and anode respectively. The CAD is denoted by d. The nanowall grows in the middle of the cathode, has width w and height h. We assume that the nanowall is metallic. Therefore, both the cathode and the nanowall is earthed and has zero electric potential. The anode is applied to a voltage of $V_0$.

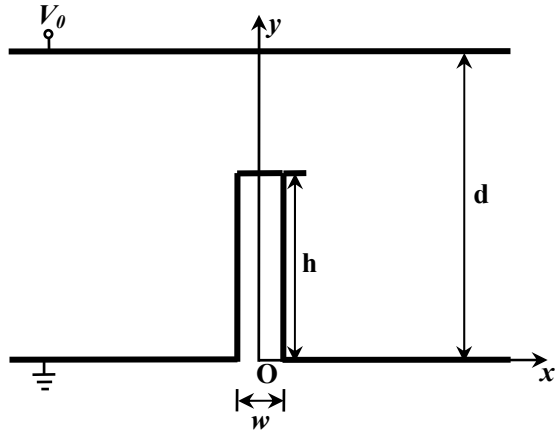

Fig. 1 The cross section view of the set up of nanowall field emission.
The nanowall is located in the middle of the bottom plane (cathode). The top plane is the anode.

The electric potential V(x,y) satisfies the two-dimensional Laplace equation and the boundary conditions that V is zero on the surfaces of cathode and nanowall and is equal to $V_0$ on the surface of the anode. The complex coordinate $z=x+iy$ will be used to specify the points of the x-y plane. Due to the left-right symmetry, actually only the first quadrant should be considered.

## 2. The conformal mapping

The solution for the electric potential can be found by two conformal transformations (CT). As shown in Fig.2, the CT from (b) to (a) maps the virtual space ($\zeta'$) to the physical space (z); the CT from (b) to (c) maps the virtual space to the target space, where the potential is easy to obtain. The representative points $W_0$, $W_{\sigma 1}$, $W_{\sigma 2}$, and $W_1$ in the physical space are corresponding to points 0, $\sigma_1$, $\sigma_2$, and 1 at the real axis of the virtual space. The left and right regions of the physical space and the corresponding regions in the mappings are indicated by two circles. The real parameters $\sigma_1$ and $\sigma_2$ will be specified in the following (eqs. (4) and (5)). We have required $0 < \sigma_1 < \sigma_2 < 1$ for the latter convenience.

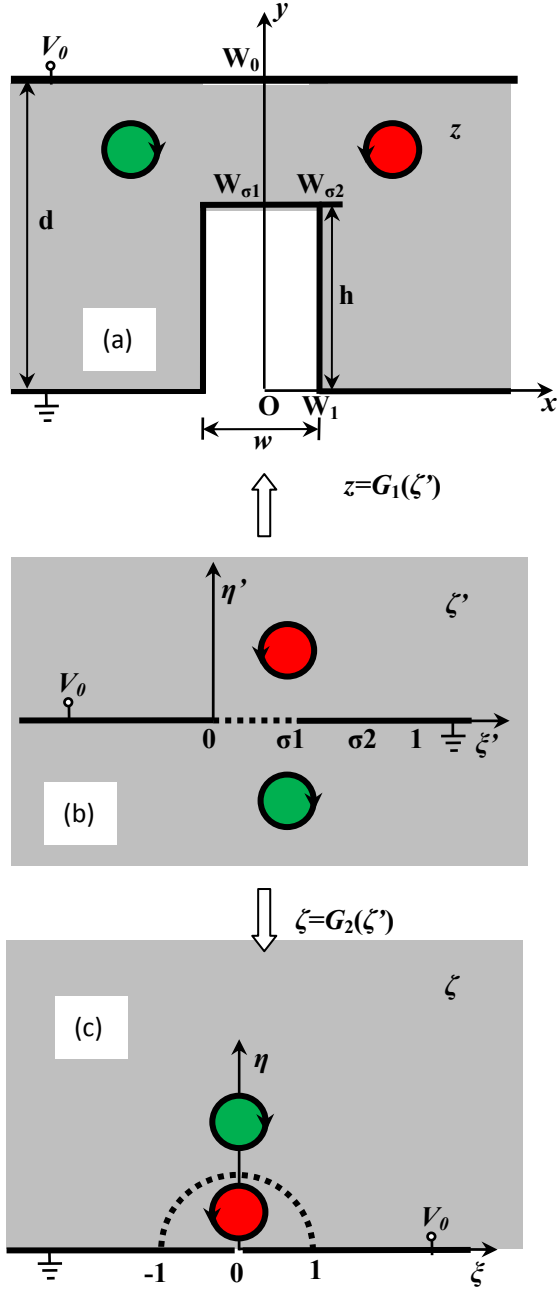

Fig.2 (color online) (a) The target plane (ζ); (b) the virtual space (η); (c) the physical space (ω).
The corresponding points are labeled by subscripts following the parameters of the target space.
Two regions are indicated by two cirles.

Denote the transformation from Fig.2(b) to Fig.2(a) by z=G$_1$(ζ'). It is given by Schwarz–Christoffel formula [15],

$$G_1(\zeta') = \frac{h}{B(1,\sigma_2)} \int_0^{\zeta'} \sqrt{\frac{\omega - \sigma_2}{\omega(\omega - \sigma_1)(\omega - 1)}} d\omega + id \qquad (1)$$

Where B(t$_1$,t$_0$) is a real functions of t$_1$ and t$_0$ defined by

$$B(t_1, t_0) = \int_{t_0}^{t_1} \sqrt{\left|\frac{t-\sigma_2}{(t-1)(t-\sigma_1)t}\right|} dt \qquad (2)$$

In this transformation, the representative points in the virtual space 0, $\sigma_1$, $\sigma_2$, and 1 are mapped to the points $W_0$, $W_{\sigma1}$, $W_{\sigma2}$, and $W_1$, respectively. The requirements of $W_0$=id, $W_{\sigma1}$=ih, $W_{\sigma2}$=r+ih, and $W_1$=r can be fulfilled if the parameters $\sigma_1$ and $\sigma_2$ satisfy the relations,

$$\frac{B(\sigma_2, \sigma_1)}{B(1, \sigma_2)} = \frac{r}{h} \qquad (3)$$

$$\frac{B(\sigma_1, 0)}{B(1, \sigma_2)} = \frac{d}{h} - 1 \qquad (4)$$

The above two equations specify the parameters $\sigma_1$ and $\sigma_2$. The functions $B(t_1,t_0)$ in (3) and (4) can be reduced to complete Elliptic integrals that only depend on the parameters $\sigma_1$ and $\sigma_2$,

$$B(1, \sigma_2) = -\frac{2(\sigma_2 - \sigma_1)}{\sqrt{(1-\sigma_1)\sigma_2}} \left[K(m) + \Pi_1(-b; m)\right] \qquad (5)$$

$$B(\sigma_2, \sigma_1) = \frac{2}{\sqrt{(1-\sigma_1)\sigma_2}} \left[\sigma_2 K(m') - \sigma_1 \Pi_1(-m'(1-\sigma_1); m')\right] \qquad (6)$$

$$B(\sigma_1, 0) = \frac{2}{\sqrt{(1-\sigma_1)\sigma_2}} \left[(\sigma_2 - 1)K(m) + \Pi_1(c_1; m)\right] \qquad (7)$$

Where we have defined $c_i = \sigma_i / (1-\sigma_i)$ for $i$=1,2; and $m = c_1 / c_2$, $m' = 1-m$, and $b = m\sigma_2 / \sigma_1$; $K(m)$ and $\Pi_1(n; m)$ are the complete elliptic integrals of the first and third kinds. The incomplete elliptic integrals of the first and third kinds, $F(\varphi|m)$ and $\Pi(\varphi|m)$ respectively, are defined in terms of the elliptic parameter $m$ and the amplitude $\varphi$ by

$$F(\varphi | m) = \int_0^{\varphi} (1 - m\sin^2 \vartheta)^{-1/2} d\vartheta \qquad (8)$$

$$\Pi(n; \varphi | m) = \int_0^{\varphi} (1 + n\sin^2 \vartheta)^{-1}(1 - m\sin^2 \vartheta)^{-1/2} d\vartheta \qquad (9)$$

At $\varphi = \pi/2$, (8) and (9) become the corresponding complete elliptic integrals, i.e., $K(m) = F(\pi/2 | m)$ and $\Pi_1(n; m) = \Pi(n; \pi/2 | m)$.

The electric potential has the most simple form in the target space (Fig.2(c)). The second conformal transformation $\zeta = G_2(\zeta')$ that maps the virtual space to the target space is

$$G_2(\zeta') = \frac{1 + \sqrt{\dfrac{\zeta'}{\zeta' - \sigma_1}}}{1 - \sqrt{\dfrac{\zeta'}{\zeta' - \sigma_1}}} \tag{10}$$

Where $\zeta'$ is the variable of the virtual space.

**3. The electric potential**

The electric potential $V(\zeta)$ in the target space (Fig.2(c)) follows the Laplace equation with the boundary condition that (i) $V(\sigma)=0$ at the negative real axis; (ii) $V(\sigma)=V_0$ at the positive real axis. The solution reads,

$$V(\zeta) = V_0\left[1 - \frac{\arg(\zeta)}{\pi}\right] \tag{11}$$

Denoting the inversed function of $G_1(\zeta')$ as $G_1^{-1}(z)$, the electric potential in the physical space has a formal expression,

$$V_{phy}(z) = V(G_2(G_1^{-1}(z))) \tag{12}$$

Actually it is difficult to use (12) directly even when $\sigma_1$ and $\sigma_2$ are known since $G_1^{-1}(z)$ is rather complicate. Fortunately, one can make a good estimation for the emission current via just knowing the potential along the y-axis because the field emission is mainly along the y-axis. The function $G_1(\zeta')$ can be expressed in elliptic integrals at the real axis of the virtual space. The point $\sigma$ in the range $0 < \sigma < \sigma_1$ of the real virtual axis is corresponding to a point at the y-axis of the physical space with $h<y<d$, $z=iy$. From (1), one has

$$y = d + \frac{h\left[(\sigma_2 - 1)F(\varphi_\sigma | m) + \Pi(c_1; \varphi_\sigma | m)\right]}{(\sigma_2 - \sigma_1)\left[K(m) + \Pi_1(-b; m)\right]} \tag{13}$$

The dependence of y on $\sigma$ is through the amplitude $\varphi_\sigma = \arcsin\left(\dfrac{\sigma}{(1-\sigma)c_1}\right)$.

To calculate the electric potential $V_{phy}(y)$ numerically, one first solves (3) and (4) for $\sigma_1$ and $\sigma_2$. Then to find the $\sigma$ for the given y via (13). Finally, from (10), (11), and (12) one obtains $V_{phy}(y)=V(G_2(\sigma))$. For instances, Fig.3 shows $V_{phy}(y)$ for r=0.3nm and h=3.6μm, with various d and a fixed $V_0/d$=50V/μm. The panels (a), (b), (c), and (d) are corresponding to d=10.μm, d=15. μm, d=20. μm, and d=25. μm, respectively. The crosses are the numerically results obtained by the FES. Results of two methods are consistent very well.

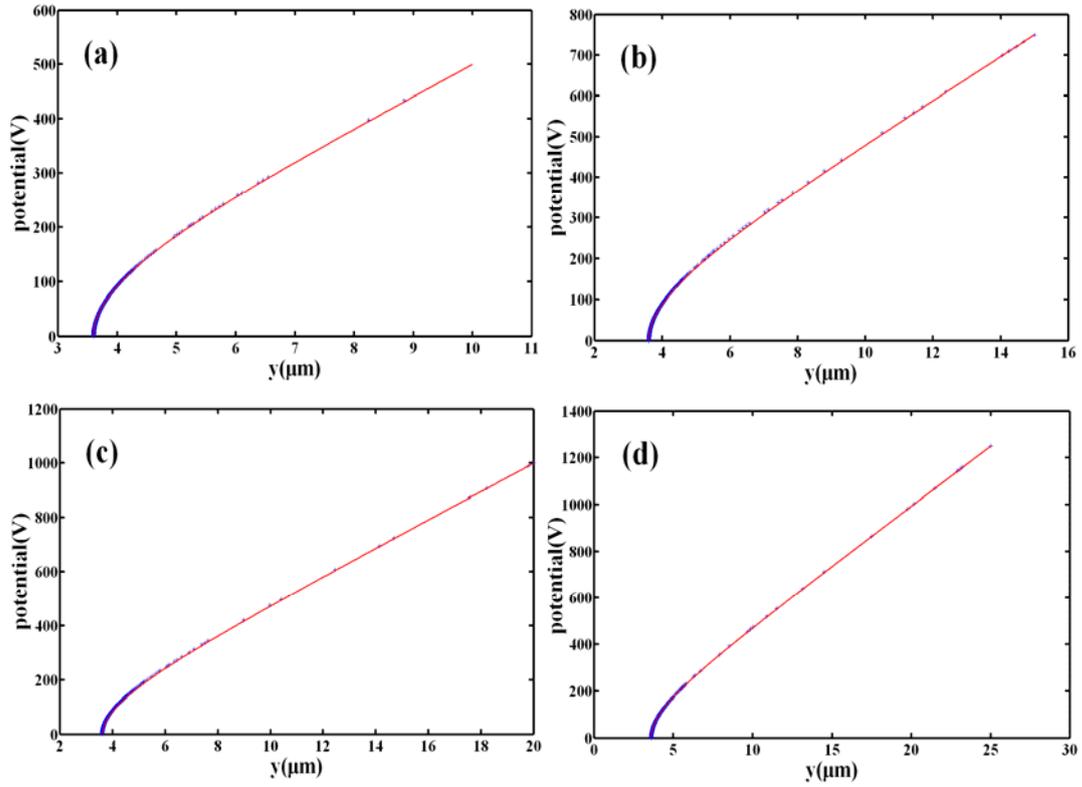

Fig.3 (color online) The electric potential along the y-axis for a nano wall of r=0.3nm and h=3.6μm. The solid(red) lines are obtained by the conformation transformation. The crosses are obtained by the finite-element simulation. (a) d=10.μm; (b) d=15. μm; (c) d=20. μm; (d) d=25. μm. The macroscopic field between the cathode and anode ($V_0/d$) is fixed as 50V/μm.

In the limit of d/h>>1, we can recover the results of [14]. In Fig.4, the finite d results of the electric potential are compared with that of the infinite d, for the same nanowall as Fig.3. The circles, triangles, squares, and crosses are the results of FES corresponding to d=10.μm, d=15. μm, d=20. μm, and d=25. μm, respectively. The solid line is the result of infinite d. Only the potential in the first 200nm from the apex is shown. One sees that the discrepancy is obvious even for d=25. μm.

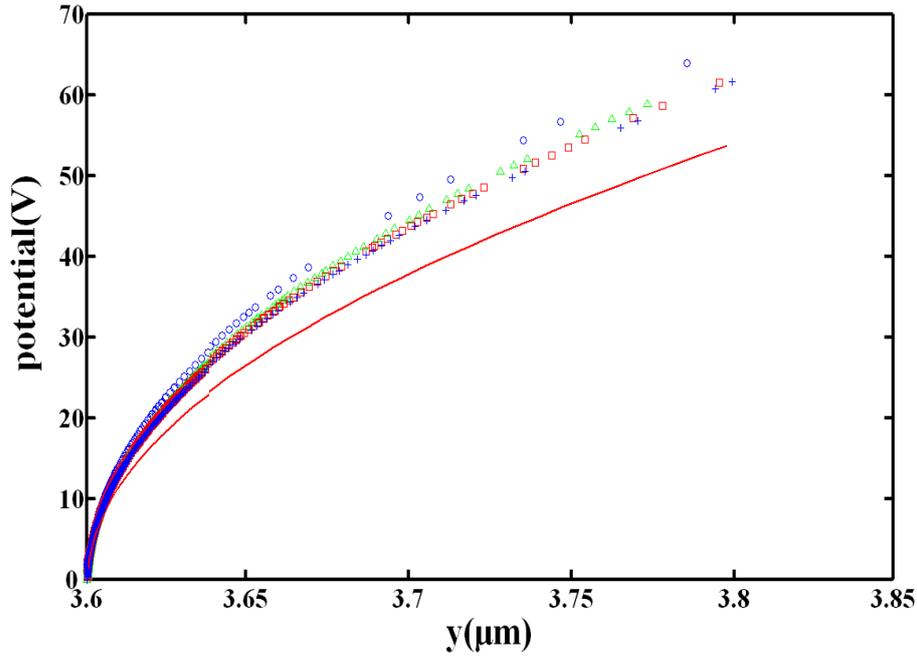

Fig.4 (color online) Comparison of the electric potentials for different cathode-anode distance (d). The y-axis is started at the apex of the nanowall that is 3.6μm in height and 2r=0.6nm in width. The circles, triangles, squares, and crosses are results obtained via the finite-element simulation, with d=10.μm, d=15. μm, d=20. μm, and d=25. μm, respectively. The solid line is the potential of infinite d ([14]).

## 4. Conclusion

The electric potential of the nanowall emitter between finite-separated cathode and anode planes has been obtained. It is a generalization of the corresponding result of infinite separation of cathode and anode [14]. Numerical results for a set of typical parameters (nanowall with 0.6nm in width, 3.6μm in height, and separations d=10.∼25. μm) are shown and compared with the finite-element simulation. It is found that the discrepancy between results of finite separation and infinite separation is large. The convergence of d approaching infinite is slow. It becomes obvious above the edge of nanowall by 20nm and higher, even for d=25. μm. In the application of electron scanner, the finite separation of cathode-anode should be considered. On the other hand, for the application of field emission, the separation is in the order of millimeters and the result of infinite separation should be used. Our results imply that one should caution to the error of artificial separation in the finite-element simulation.

ACKNOWLEDGMENTS

The project is supported by the National Natural Science Foundation of China (Grant Nos. 10674182, 10874249, and 90306016) and the National Basic Research Programme of China (2007CB935500).